\documentclass{arxiv}
\usepackage{amssymb}
\usepackage{epsfig}
\usepackage{subfigure}
\usepackage{cite}
\usepackage{rotating}
\usepackage{caption}

\begin{document}

\catchline{}{}{}{}{}
%%%%%%%%%%%%%%%%%%%%%%%%%%%%%%%%%%%%%%%%%%%%%%%%%%%%%%%%%%%%%%%%%%%

\title{The Benford law behavior of the religious activity data
}

\author{T. A. Mir$^{*}$}

\address{Nuclear Research Laboratory, Astrophysical Sciences Division,\\ Bhabha Atomic Research Centre,\\  Srinagar-190 006, Jammu and Kashmir, India\\
$^{*}$taarik.mir@gmail.com}

\maketitle

%\pub{Received (Day Month Year)}{Revised (Day Month Year)}

\section{Abstract}

An important aspect of religious association is that adherents, as part of their religious duty, carry out various activities. One religious group known for keeping the elaborate records of day-to-day activities of its members is the Jehovah's Witnesses (JWs)-a worldwide Christian religious group. We analyze the historical records of the country-wide data associated with twelve different religious activities of JWs to see if there are any patterns in the distribution of the first digits as predicted by Benford's law. This law states that the first digits of numbers in data sets are not uniformly distributed but often, not always, follow a logarithmic distribution such that the numbers beginning with smaller digits appear more frequently than those with larger ones. We find that the data on religious activities like peak publishers, pioneer publishers, baptizations, public meetings, congregations, bible studies, time spent in door-to-door contacts, attendances at the memorial services, total literature and individual magazines distributed, new subscriptions and back-calls received excellently conforms to Benford's law. 
\section{Keywords}

Benford's law; Jehovah's Witnesses; religious activities 
%\end{abstract}

%\ccode{PACS Nos.: include PACS Nos.}

\section{Introduction}	

Intuitively one might guess the occurrence of numbers in data sets from different sources to be uniform with each digit from 1 to 9 having a proportion of appearance of about 11\% as the first digit of numbers, independent of its magnitude. But surprisingly according to Benford's law the numbers in data sets from many real world phenomena appear such that the numbers with smaller initial digits occur more frequently than those beginning with larger digits\cite{Benford}. 
The peculiar phenomenon first reported by Simon Newcomb following his observation that the initial pages of the logarithmic table books were dirtier than the later ones was rediscovered by F. Benford through a similar observation\cite{Newcomb}. Newcomb merely reported his finding and did not follow on it. The phenomenon attained much popularity only after Benford (hence the name Benford's law) tested the accuracy of this observation  on the large volume of data he collected from diverse fields e.g. the physical constants, the atomic and molecular masses, street addresses, the length of rivers etc. and concluded that the occurrence of first significant digits follow a logarithmic distribution\cite{Benford}. \\ 
\begin{equation}
P(d)= log_{10}(1+\frac{1}{d}), d= 1, 2, 3...,9
%p=\frac{e^{-b.\delta t}(b.\delta t)^n}{n!}
\end{equation}
where P(d) is the probability of a number having the first non-zero digit d and $log_{10}$ is logarithmic to base 10\\
According to equation (1), in a given data set the probability of occurrence of a certain digit as first significant digit decreases logarithmically as the value of the digit increases from 1 to 9. The theoretical proportions for each of the digits from 1 to 9 to be first significant digit are as shown in Table 1.
\begin{table}[h]
\tbl{The distribution of first significant digits as predicted by Benford's law}
{\begin{tabular}{@{}ccccccccccc@{}} \toprule
Digit \hphantom{00} & 1 & 2 & 3 & 4 & 5 & 6 & 7 & 8 & 9 \\
Proportion \hphantom{00} & 0.301 & 0.176 & 0.125 & 0.097 & 0.079 & 0.067 & 0.058 & 0.051 & 0.046 \\ \botrule
\end{tabular} \label{ta1}}
\end{table} 
\newline
Though the mathematical form of Benford's law (eq. 1) is quite simple, a complete explanation is still lacking\cite{Berger}. Nevertheless, the two key properties of the law i.e. scale-invariance and base-invariance have been explained\cite{Hill, Hill1}. The number of fields where the presence of the law has been reported is too large to be listed here. For a comprehensive information on the numerous data sets which obey Benford's law and its applications across multiple disciplines the reader is referred to ``Benford Online Bibliography''\cite{Online}. Some examples are the demographic data\cite{Sandron, Mir}, financial data\cite{Pietronero, Giles, Mir1}, scientometric data\cite{Campanario, Egghe, Mir2} and scientific data\cite{Pain, Shao, Sambridge}. Practically Benford's law is used as a statistical sleuth to detect data manipulation. Any departure from the law is taken as an indication of the presence of inconsistencies in the data under investigation. The law has been used to detect anomalies in survey data from social processes\cite{Judge}, manipulated data submitted by the companies for tax evasion\cite{Nigrini}, enhanced figures in macroeconomic data reported by countries for strategic advantage\cite{Michalski, Rauch} and for uncovering fraud in election vote counting and campaign finances\cite{Mebane, Cho}. In the same line of thought the law has been used to investigate the reliability of yearly financial reports on income and expenses of the Antoinist community\cite{Clippe}.
\newline
One of the key social attributes of the human beings is their association with religions. The prevalence of Benford's law has been tested for the numbers appearing in the religious texts  like the Book of Mormons\cite{Lindsay}, the Quran\cite{Motahari} and the Bible\cite{Makous, Salsburg}. Further, the law has been used to assess the reliability of the religious demography data by exploring the conformity of country-wide data on the number of adherents of the world's seven major religions and the adherent data on all the religions, except Christianity, has been found to conform to the law\cite{Mir}. Furthermore, the significant digit distribution of the three major Christian denominations, Catholicism, Protestantism and Orthodoxy, is found to obey Benford's law.
\newline
By analyzing the adherent distribution of largest religions the study in Ref.\cite{Mir} took a global view of religious association. However, on a more local level there are smaller religious groups, sects and cults which are the breakaway off shoots of the dominant religions. Being at odds with the norms established by the dominant religions the social, political and economic conduct of such smaller religious groups is controversial, often are being accused of wrongdoings\cite{Beckford}. One wonders whether the supposed misconduct of these sects is also reflected in the data on their day-to-day activities and the answer to this pertinent question is to be found in the official data, if there are any, of such religious groups. Fortunately, there is one religious group called Jehovah's Witnesses (JWs) whose official reports are in public domain\cite{Witnesses, Wah}. JWs is a worldwide Christian religious group which despite its controversial stance on a range of issue like refusal to (i) accept blood transfusion (ii) take part in wars and (iii) honor the national flags of countries they live in, has seen an extraordinary growth over the years\cite{Stark}. Witnesses are known for their active door-to-door proselytism and all practicing members actively involved in a public preaching work are required to keep elaborate records of their religious activities. The reports of members from individual countries generate enormous statistics which after compilation into comprehensive data tables are published in the yearbook of JWs\cite{Wah}. The yearbook for a particular year details the statistics on religious activities carried out during the preceding year called the service report year e.g. the 2013 yearbook reports the data on the activities carried out during the year 2012. 
\newline
In the present study the author analyzed the historical records of the data associated with different activities of JWs to see if there are any patterns in the occurrence of digits as predicted by Benford's law. Such a study would be a natural extension of studies showing the prevalence of Benford's law for the adherent distribution of major world religions and for numbers in religious texts. We focus on twelve distinct measures of religious activity of JWs(i) total number of publishers (ii) monthly average of full-time publishers (pioneers) (iii) total time in hours spent in the public preaching work (iv) number of Bible studies conducted (v) attendance at the memorial services (vi) number of people baptized (vii) congregations held (viii) public meetings (ix) total literature distributed (x) new subscriptions received (xi) individual magazines distributed and (xii) back-calls received. We find that country-wide data on all these religious activities follow Benford's law.

\section{Data}
According to 2013 yearbook Witnesses are found world over with their existence being reported in 239 lands\cite{yb13}. However, the individual data on their religious activities are published for only 209 countries. In addition the statistics are also published collectively for 30 countries where the activities of the JWs are banned and their operation is underground. We analyse the service reports of Witnesses available online from the official website of the group for a 12-year period from 2001 to 2012\cite{Witnesses}. The data for 12-year period from 1947 to 1958 was obtained from Ref.\cite{jwfacts}. 

\subsection{Data analysis and Results}

Though JWs were formed in the 1870s, the detailed statistical data was tabulated only after the year 1946 and prior to that the data was reported separately for the lands in which their members were operating. Over the years the number of individual lands from which the data is reported has  increased significantly. This is firstly due to Witnesses having won legal status in many countries and secondly due to reports now being filed from individual countries which earlier were the members of the erstwhile USSR, where Witnesses were banned\cite{Dirksen}. Thus legal, political and social attitude towards Witnesses has changed dramatically over the years from being hostile around the period of World War II to having gained more acceptability in recent years. It would be interesting to check whether the dramatic change in attitude towards Witnesses has had any impact on the quality of their yearly reports. We obtained the relevant data for the period of years 2001-2012 from the official website of the JWs\cite{Witnesses} and for period 1947-1958 from Ref.\cite{jwfacts}. The two periods of data are so chosen that our analysis captures the degree of quality of data from periods of divergent attitudes towards Witnesses and also at the same takes into account the increase in the number of lands from which the data are reported. Benford's law works better for larger data sets is well known\cite{Nigrini2}. As a representative of these two time periods we show the detailed statistical analysis of the yearbooks of JWs, one from each of the two periods, of the country-wide data of their religious activities in Tables 1 and 2. In Table 1 the data on the six religious activities is taken from the service report of year 2002. The $N_{Obs}$, the number of counts appearing in the corresponding data set as first significant digit, against each digit from 1 to 9 are shown for each activity in columns 2, 3, 4, 5, 6 of Tables 1. We also show $N_{Ben}$, the corresponding counts (in brackets) for each digit as predicted by Benford's law: 
\begin{equation}
N_{Ben}= N log_{10}(1+\frac{1}{d})
%p=\frac{e^{-b.\delta t}(b.\delta t)^n}{n!}
\end{equation} 
along with the root mean square error ($\Delta{N}$) calculated from the binomial distribution\cite{Shao}
\begin{equation}
\Delta{N}= \sqrt{NP(d)(1-P(d))}
%p=\frac{e^{-b.\delta t}(b.\delta t)^n}{n!}
\end{equation} 
where $N$ for each religious activity is the total number of records i.e. the number of countries the data are reported from. 
For example, as shown in column 2 of Table 1, the number of peak publishers are reported from 206 lands. The observed count for digit 1 as first significant digit is 70 whereas the expected count from Benford's law is 62 with an error of about 6.6. Note that since number of records for first four activities is same, the expected count from Benford's law and the corresponding error are also same and are shown only once in Column 5. From a visual inspection of Table 2 it is found that the observed and expected digit distributions are in reasonable agreement. This is further illustrated in sub figures a-f of Fig. 1 where the observed proportion of the first digits with those expected from Benford's law are compared. Similar pattern of the decrease in proportion of the digits with increase in value of the respective digits can be easily seen in all six cases. However, having a distribution of digits exactly same as a Benford one is practically impossible\cite{Nigrini2} and one needs to quantify the level of agreement of the observed and expected digit distributions. We use Pearson's $\chi^{2}$ test to examine the goodness of fit.
\begin{equation}
  \chi^{2}(n-1) =\sum_{i=1}^n\dfrac{(N_{Obs}-N_{Ben})^{2}}{N_{Ben}}
%p=\frac{e^{-b.\delta t}(b.\delta t)^n}{n!}
\end{equation}

In our case $n=9$ which means we have $n-1=8$ degrees of freedom. Under $95\%$ confidence level (CL) the critical value for acceptance or rejection of null hypothesis - the observed and theoretically predicted digit distribution are same, is $\chi^{2}(8)$=15.507. If the value of the calculated $\chi^{2}$ is less than the critical value then we accept the null hypothesis and conclude that the data fits Benford's law.
\newline
For the country-wide number of peak publishers (column 2 of Table 1), the $\chi^{2}$= 13.990 (the last row and column 2 of Table 1) is smaller than $\chi^{2}(8)=15.507$ and thus the null hypothesis is accepted indicating in turn that the data on the country-wide distribution of peak publishers follows Benford's law. Further we also studied the digit distribution of data on country-wise number of hours spent in door-to-door campaign, number of Bible studies conducted, number of attendances at memorial services, number of people baptized, number of average pioneer publishers and number of congregations held which are shown in Columns 3-8. The respective $\chi^{2}$ values are 8.944, 8.130, 13.970, 9.118, 4.880 and 9.677, all lower than $\chi^{2}(8)=15.507$ and hence null hypothesis must be accepted. 
\newline 
An integral part of proselytism of the JWs is the distribution of literature. In order to convince a prospective convert during their door-to-door campaign, Witnesses first hand him a copy of one of their magazines and subsequently if the prospect shows some interest and is willing to speak with them, they return in a week or so for more discussions, called a "back-call". The publishing houses of JWs prints literature in more than 180 languages which is distributed across 235 countries\cite{WT}. However, the recent yearbooks of JWs do not detail the information on the country-wide distribution of literature\cite{yb13}. To look for the relevant information the author while going through the historical records of the JWs found that besides information on literature the older yearbooks also contain data on some other religious activities like public meetings and back-calls received, which are not reported now. For illustration we chose to analyze statistics reported in the 1959 Yearbook i.e. service report for year 1958. This report has data on five other religious activities besides those analyzed in Table 1. Particularly the report has data on total literature distributed, publication of individual magazines and new subscriptions received.  The analysis is detailed in Table 2. Clearly $\chi^{2}$'s for all the five religious activities are less than the critical value of 15.507. The smaller values of Pearson's $\chi^{2}$ make the null hypothesis acceptable and hence Benford's law describes the distribution of significant digits in each case (see Table 3).
\newline
We summarize the results of our first digit analysis of the JWs data for the years from 1947 to 1958 in Table 4 and for the years from 2001 to 2012 in Table 5. For each year the sample size i.e. the number of records of each religious activity are shown along with the $\chi^{2}$ which is shown in brackets. For example in Table 4, for the report year 2001 (Column 1) the data on peak publishers is reported from 206 territories (Column 2) and the respective $\chi^{2}$ of 12.369 is shown alongside in brackets. The $\chi^{2}$ for the samples of activities which deviate from the Benford law are shown in boldface. As can be seen the digit distribution of an overwhelming majority of the samples follows Benford's law (see Fig. 2).    
\newpage
\begin{table}
\tbl{The significant digit distribution of country-wide religious activities of JW for service report year 2002}
{\begin{tabular}{@{}llllllll@{}} \toprule
First &  Peak & Hours & Bible & Memorial  & Baptizations & Congregations \\
digit &  publishers & spent & studies & service atten-&  & \\
 &  (206) & (206) & (206) & dances (206) & (191) &  (204) \\ \colrule
$1$ \hphantom{00} &  79  & 60 & 65 & 51 (62.0$\pm$6.6) & 53 (57.5$\pm$6.3) &  65 (61.4$\pm$6.6) \\\\

$2$ \hphantom{00} & 39 & 44  & 35 & 43 (36.3$\pm$5.5) & 29 (33.6$\pm$5.3) &  45 (35.9$\pm$5.4) \\\\

$3$ \hphantom{00} &  26 & 32 & 22 & 26 (25.7$\pm$4.8) & 31 (23.9$\pm$4.6) &  34 (25.5$\pm$4.7) \\\\

$4$ \hphantom{00} & 11 & 17  & 19 & 28 (20.0$\pm$4.3) & 24 (18.5$\pm$4.1) & 13 (19.8$\pm$4.2) \\\\ 

$5$ \hphantom{00} & 12 & 17 & 18 & 16  (16.3$\pm$3.9) & 12 (15.1$\pm$3.7) & 9 (16.2$\pm$3.9) \\\\

$6$ \hphantom{00} & 16 & 15 & 18 & 15 (13.8$\pm$3.6) & 15 (12.8$\pm$3.5) &  10 (13.7$\pm$3.6) \\\\

$7$ \hphantom{00} & 6 & 10 & 7 & 8  (12.0$\pm$3.4) & 14 (11.1$\pm$3.2) & 14 (11.8$\pm$3.3) \\\\

$8$ \hphantom{00} & 10 & 8 & 13 & 13 (10.5$\pm$3.2) & 5 (9.8$\pm$3.1) &  8 (10.4$\pm$3.2) \\\\

$9$ \hphantom{00} & 7 & 3 & 9 & 6 (9.4$\pm$3.0) & 8 (8.7$\pm$2.9) &  6 (9.3$\pm$3.0) \\ 
\botrule
\bf$\chi^{2}$ \hphantom{00} & \bf13.990 & \bf8.944 & \bf8.130  & \bf13.970 & \bf9.118 &  \bf9.677 \\ \botrule
\end{tabular} \label{ta1}}
%\end{table} 
\end{table}

\begin{table}
\tbl{The significant digit distribution of country-wide religious activities of JW for service report year 1958}
{\begin{tabular}{@{}lllllll@{}} \toprule
First & Pioneer & Public & Total & New & Individual & Back-calls \\
digit & publishers & meetings & literature & subscriptions & Magazines & \\
 & (147) & (145) & (167) & (153) & (162) & (167) \\ \colrule
$1$ \hphantom{00} & 42 (44.3$\pm$5.6) & 51 (43.7$\pm$5.5) &  50 (50.3$\pm$5.9) & 46 (46.1$\pm$5.7) & 50 (48.8$\pm$5.8) & 49 (50.3$\pm$5.9) \\\\

$2$ \hphantom{00} &  34 (25.9$\pm$4.6) & 21 (25.3$\pm$4.6) &  29 (29.4$\pm$4.9) & 24 (26.9$\pm$4.7) & 27 (28.5$\pm$4.9) & 37 (29.4$\pm$4.9) \\\\

$3$ \hphantom{00} & 12 (18.4$\pm$4.0) & 14 (18.1$\pm$4.0) & 21 (20.9$\pm$4.3) & 17 (19.1$\pm$4.1) & 23 (20.2$\pm$4.2) & 16 (20.9$\pm$4.3) \\\\

$4$ \hphantom{00} & 11 (14.3$\pm$3.6) & 15 (14.1$\pm$3.6) & 18 (16.2$\pm$3.8) & 11 (14.8$\pm$3.7) & 20 (15.7$\pm$3.8) & 20 (16.2$\pm$3.8) \\\\ 

$5$ \hphantom{00} & 16 (11.6$\pm$3.3) & 10 (11.5$\pm$3.3) & 12 (13.2$\pm$3.5) & 16 (12.1$\pm$3.3) & 6 (12.8$\pm$3.4) & 12  (13.2$\pm$3.5) \\\\

$6$ \hphantom{00} & 10 (9.8$\pm$3.0) & 12 (9.7$\pm$3.0) & 9 (11.2$\pm$3.2) & 8 (10.2$\pm$3.1) & 7 (10.9$\pm$3.2) & 8 (11.2$\pm$3.2) \\\\

$7$ \hphantom{00} & 10 (8.5$\pm$2.8) & 6 (8.4$\pm$2.8)  & 7 (9.7$\pm$3.0) & 13 (8.9$\pm$2.9)  & 16 (9.4$\pm$3.0)  & 9  (9.7$\pm$3.0) \\\\

$8$ \hphantom{00} & 8 (7.5$\pm$2.7) & 9 (7.4$\pm$2.7) & 12  (8.5$\pm$2.9) & 9  (7.8$\pm$2.7) & 7 (8.3$\pm$2.8) & 7  (8.5$\pm$2.9) \\\\

$9$ \hphantom{00} & 4 (6.7$\pm$2.5)  & 7 (6.6$\pm$2.5) & 9 (7.6$\pm$2.7) & 9 (7.0$\pm$2.6) & 6 (7.4$\pm$2.7) & 9 (7.6$\pm$2.7) \\ 
\botrule
\bf$\chi^{2}$ \hphantom{00} & \bf8.632 & \bf4.822 &  \bf3.135 & \bf5.948 & \bf11.778 & \bf5.613  \\ \botrule
\end{tabular} \label{ta1}}
\end{table}

\begin{figure}
\begin{center}
\begin{minipage}[b]{.9\linewidth}
\vspace*{-5pt}
\hspace*{20pt}
\centering
\begin{tabular}{cc}
\hspace*{-50pt}
\vspace*{0pt}
\subfigure[Peak publishers]{\label{fig:edge-a}\includegraphics[width=0.55\linewidth, height=0.5\linewidth,  clip=]{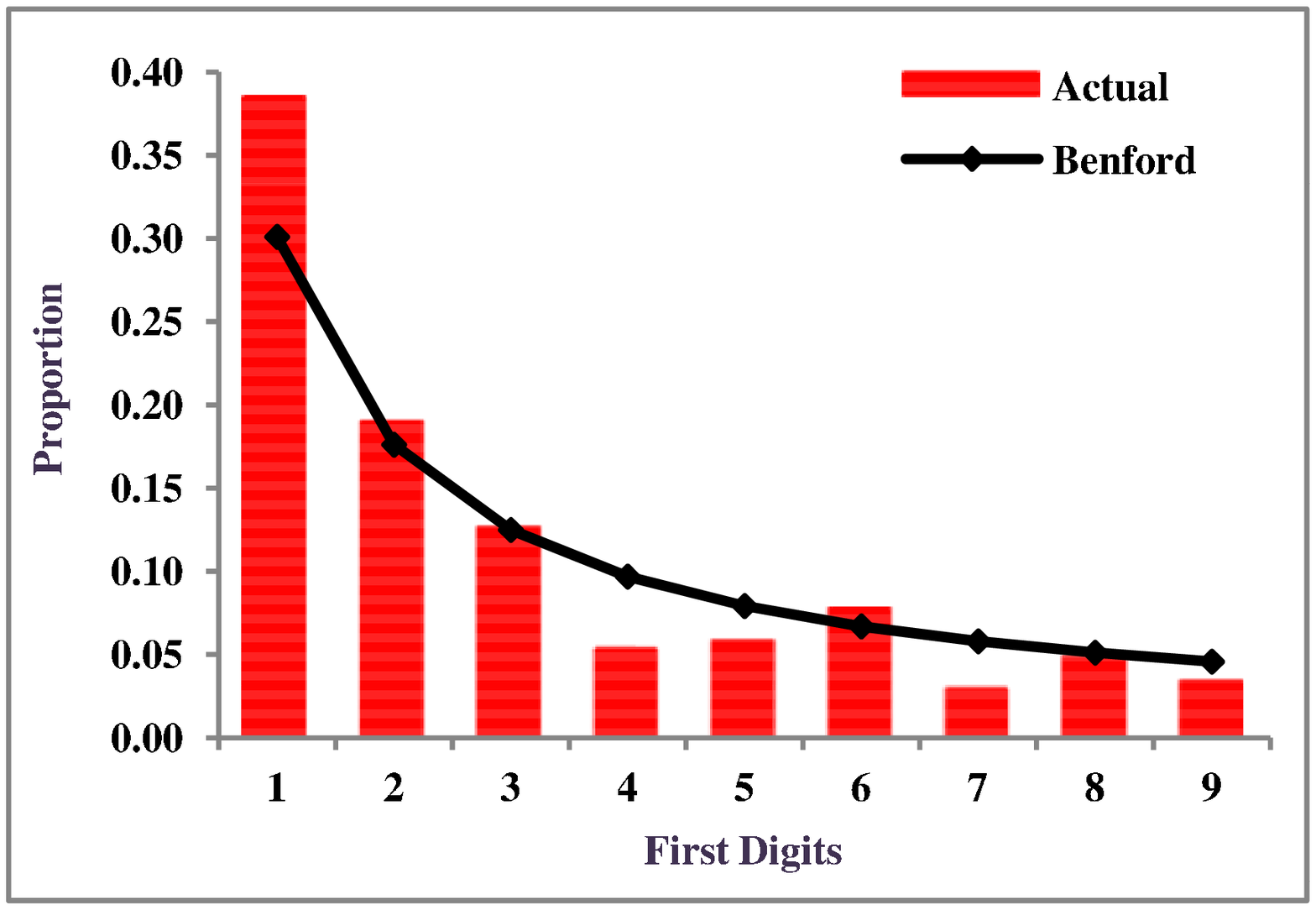}}
\hspace*{-5pt}
\subfigure[Congregations]{\label{fig:edge-b}\includegraphics[width=0.55\linewidth, height=0.5\linewidth,  clip=]{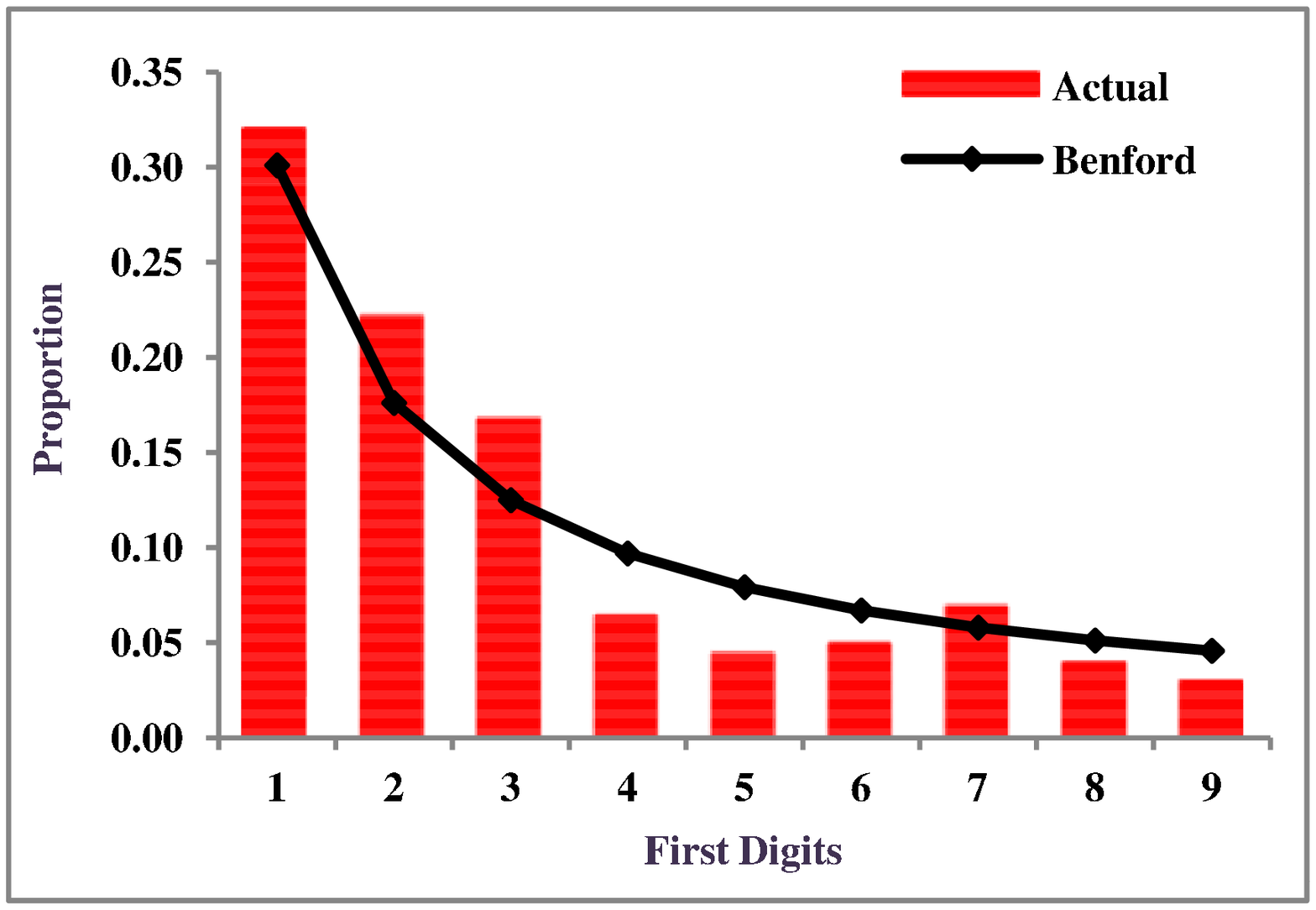}}\\
\hspace*{-50pt}
\vspace*{10pt}
\subfigure[Hours]{\label{fig:edge-a}\includegraphics[width=0.55\linewidth, height=0.5\linewidth, clip=]{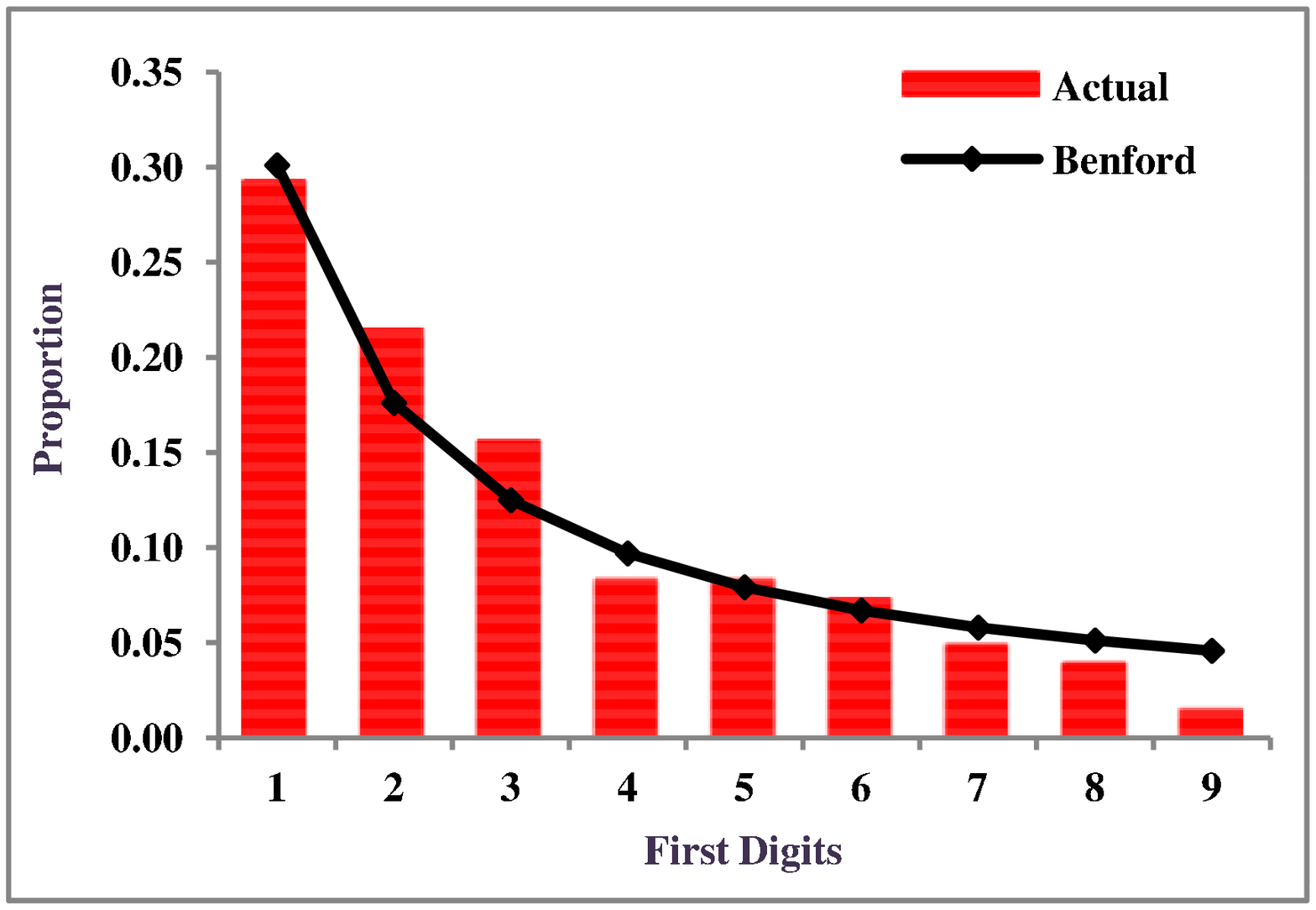}}
\hspace*{-5pt}
\subfigure[Bible studies]{\label{fig:edge-b}\includegraphics[width=0.55\linewidth, height=0.5\linewidth, clip=]{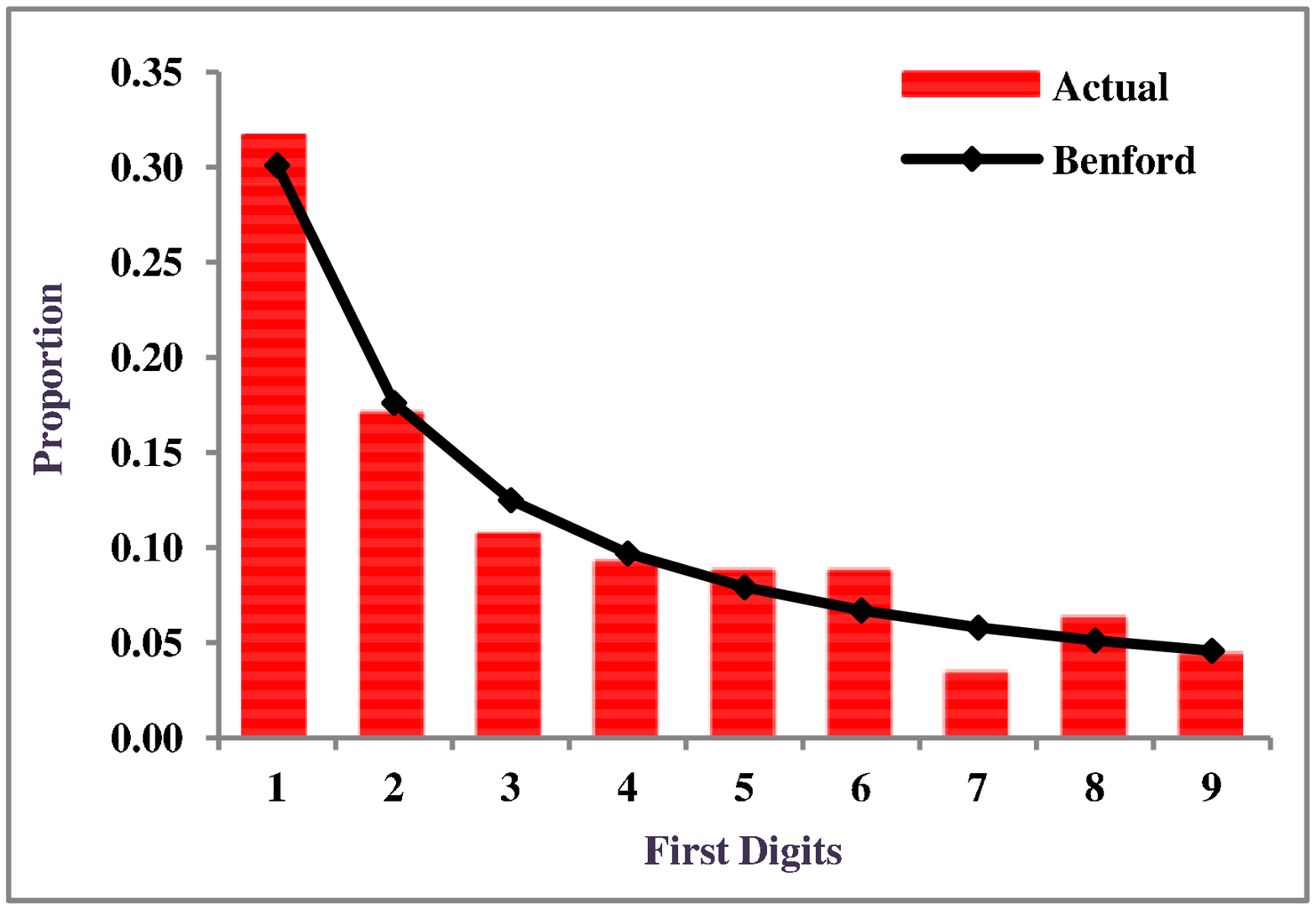}}\\
\hspace*{-50pt}
\vspace*{-150pt}
\subfigure[Memorial service attendance]{\label{fig:edge-a}\includegraphics[width=0.55\linewidth, height=0.5\linewidth, clip=]{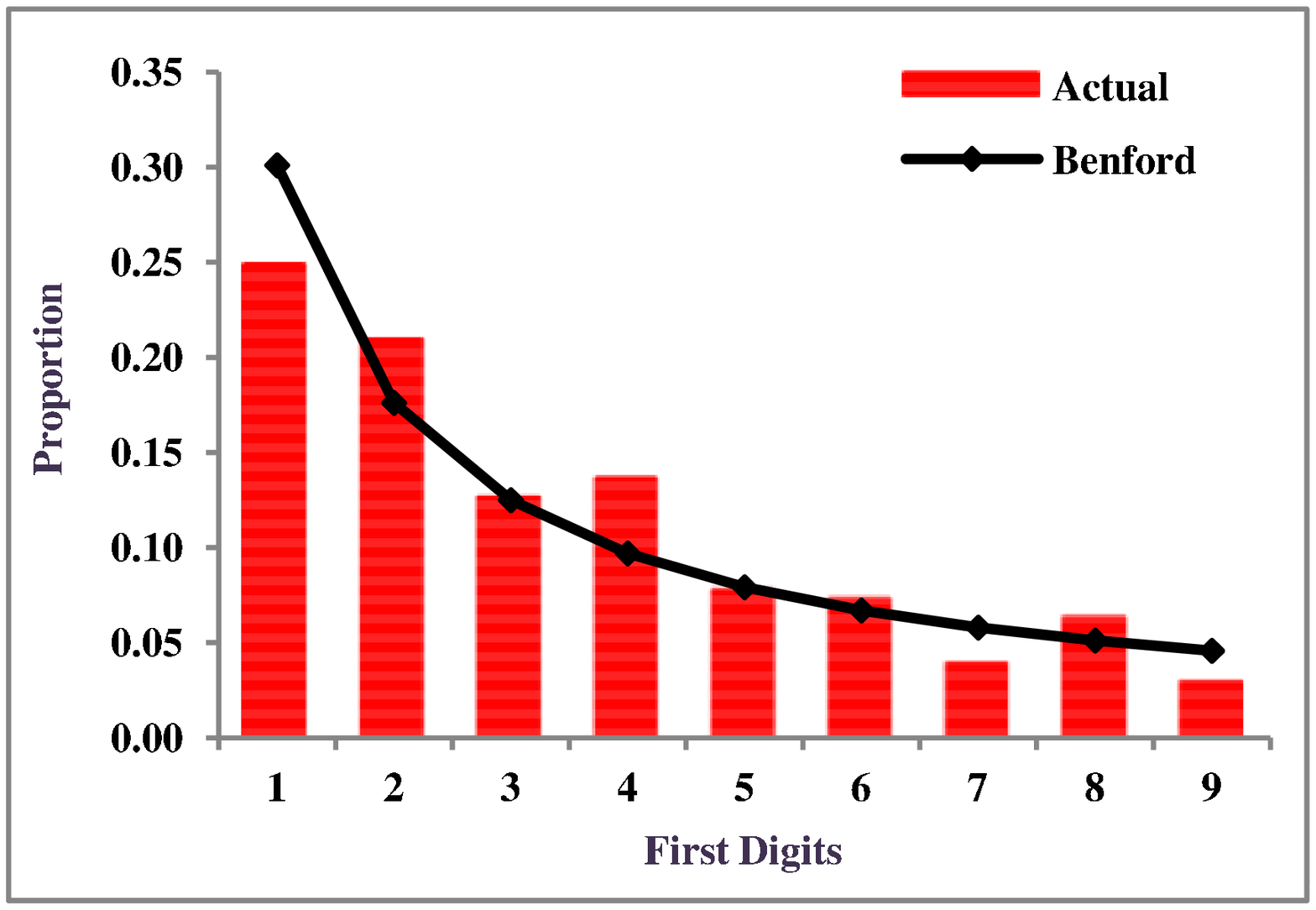}}
\hspace*{-5pt}
\subfigure[Baptizations]{\label{fig:edge-b}\includegraphics[width=0.55\linewidth, height=0.5\linewidth, clip=]{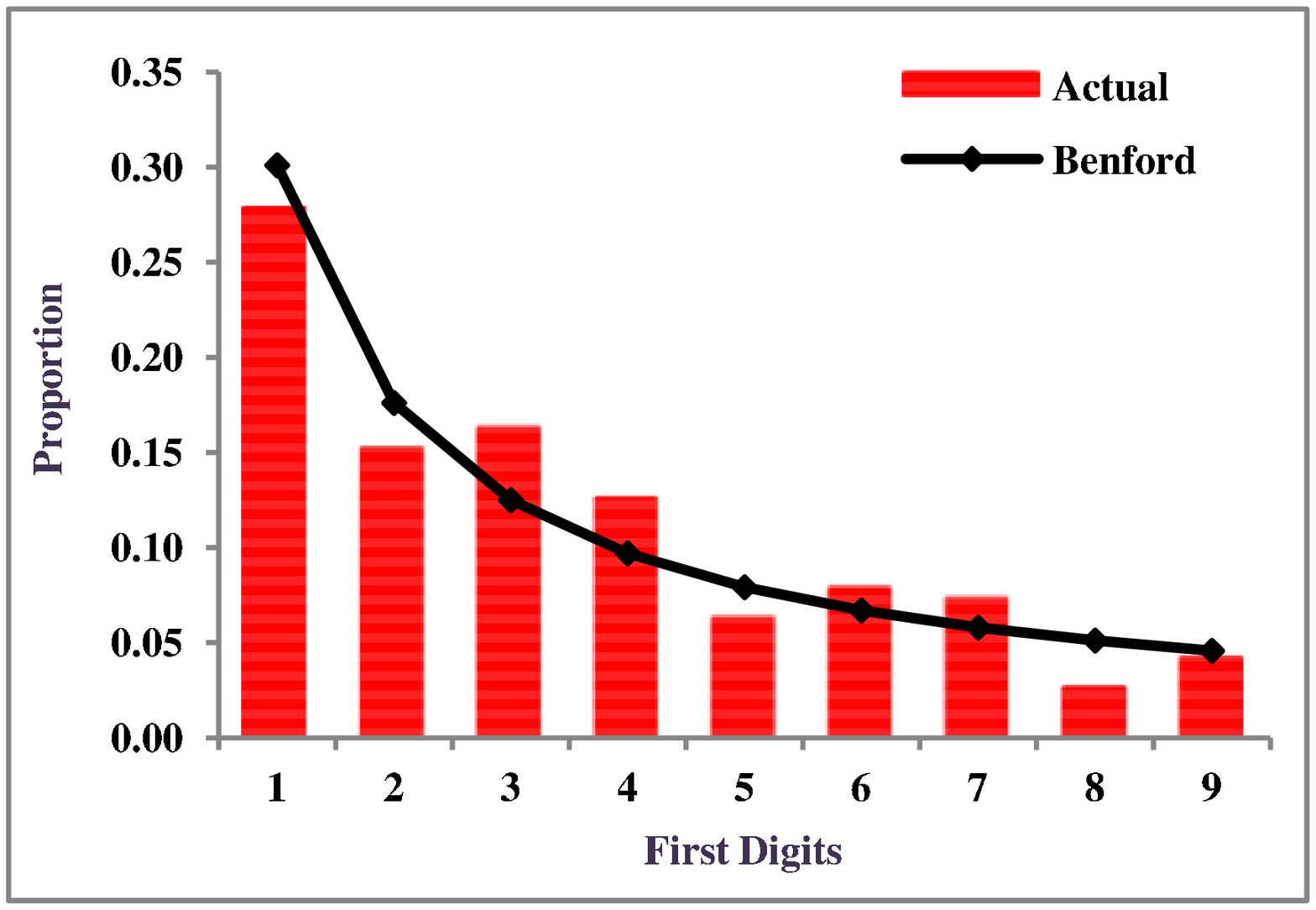}}\\
\hspace*{-74pt}
\vspace*{110pt}
\hspace*{20pt}
\vspace*{-70pt}
\end{tabular}
\vspace*{90pt}
\hspace*{20pt}
\end{minipage}
\end{center}
\caption{Observed and Benford distributions of significant digits for country-wide religious activity data of JW for service report year 2002}
\label{fig:edge}
\end{figure}

\begin{figure}
\begin{center}
\begin{minipage}[b]{.9\linewidth}
\vspace*{-5pt}
\hspace*{20pt}
\centering
\begin{tabular}{cc}
\hspace*{-50pt}
\vspace*{0pt}
\subfigure[Pioneer publishers]{\label{fig:edge-a}\includegraphics[width=0.55\linewidth, height=0.5\linewidth,  clip=]{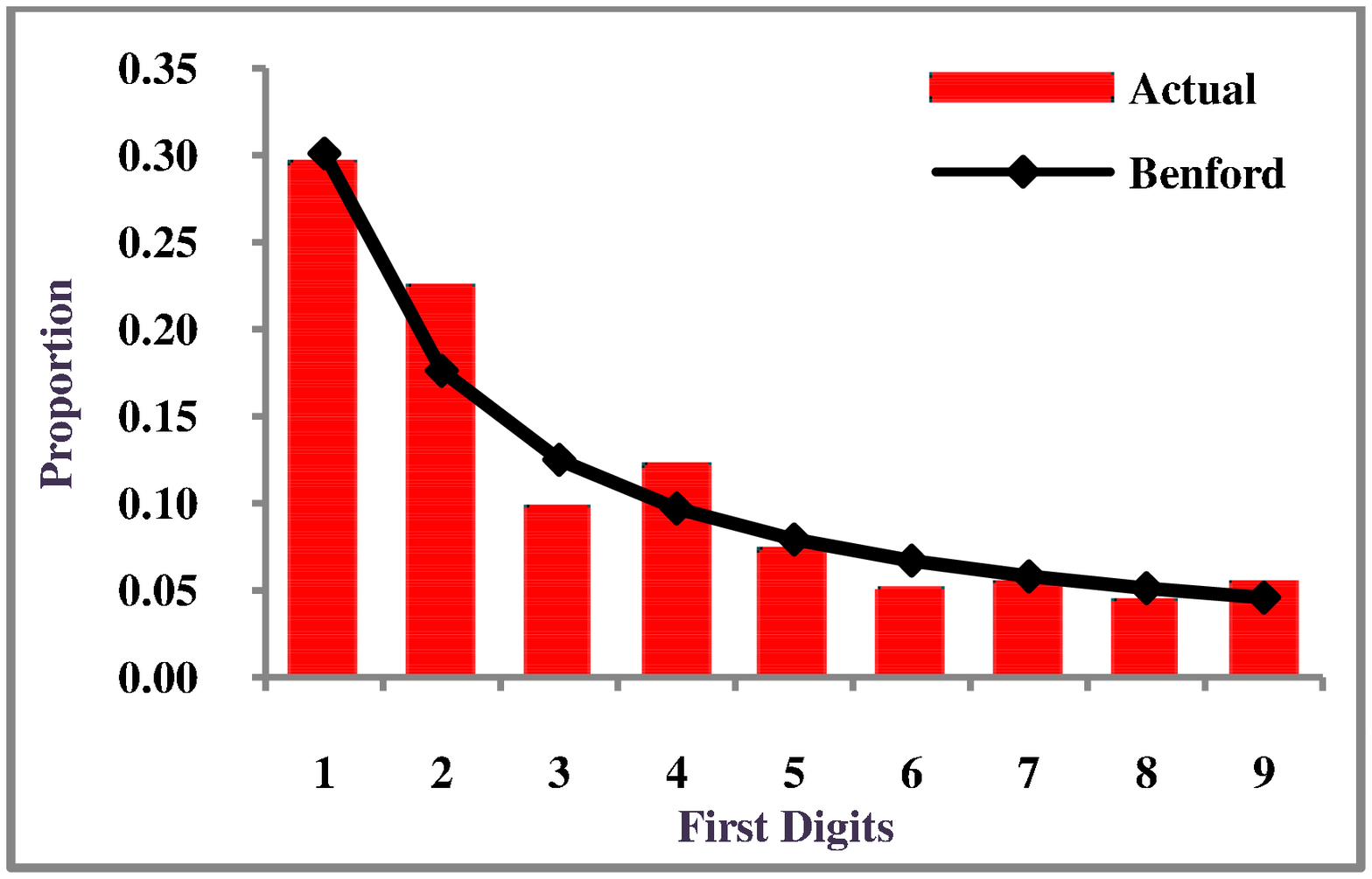}}
\hspace*{-5pt}
\subfigure[Public meetings]{\label{fig:edge-b}\includegraphics[width=0.55\linewidth, height=0.5\linewidth,  clip=]{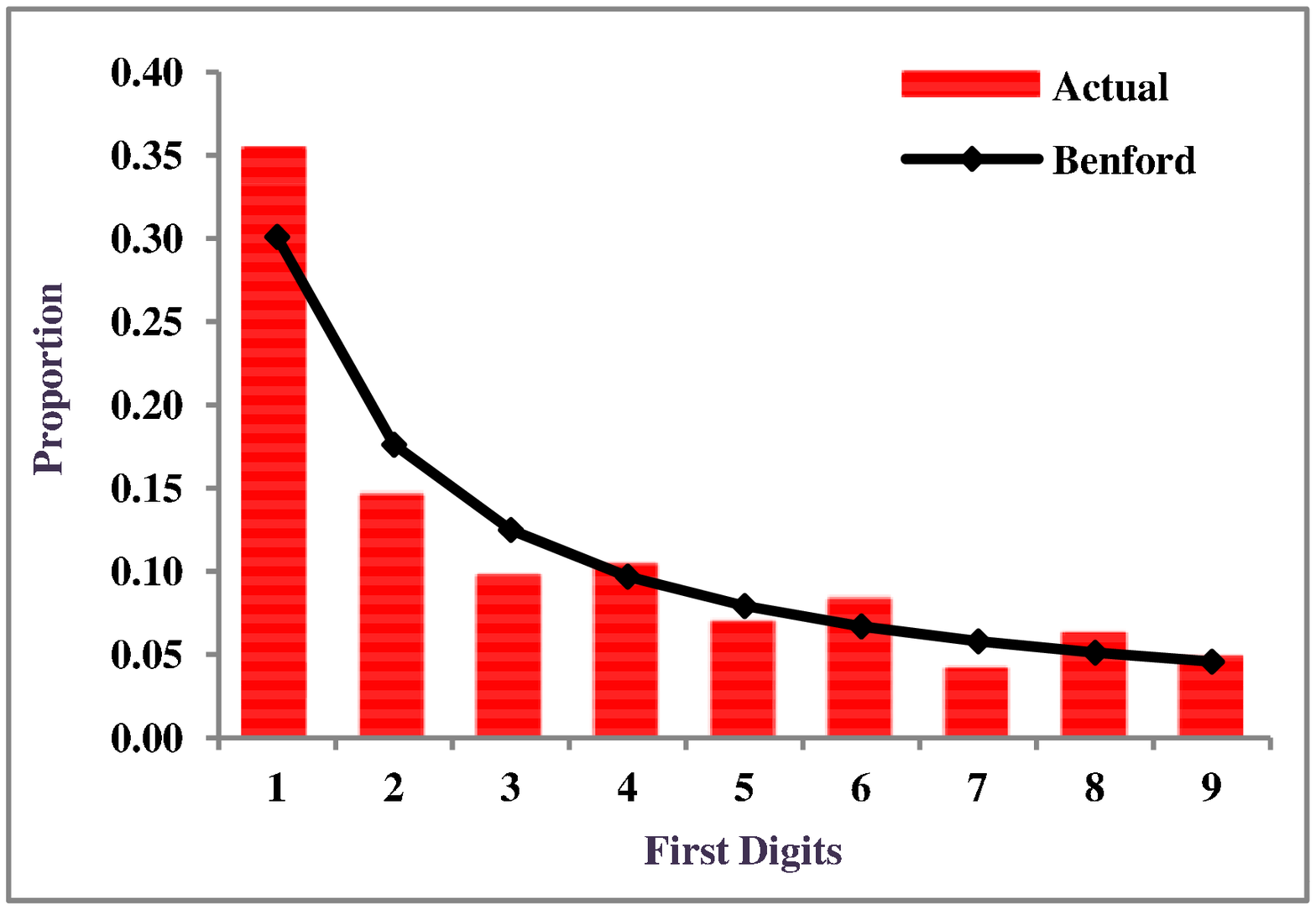}}\\
\hspace*{-50pt}
\vspace*{10pt}
\subfigure[Total literature]{\label{fig:edge-a}\includegraphics[width=0.55\linewidth, height=0.5\linewidth, clip=]{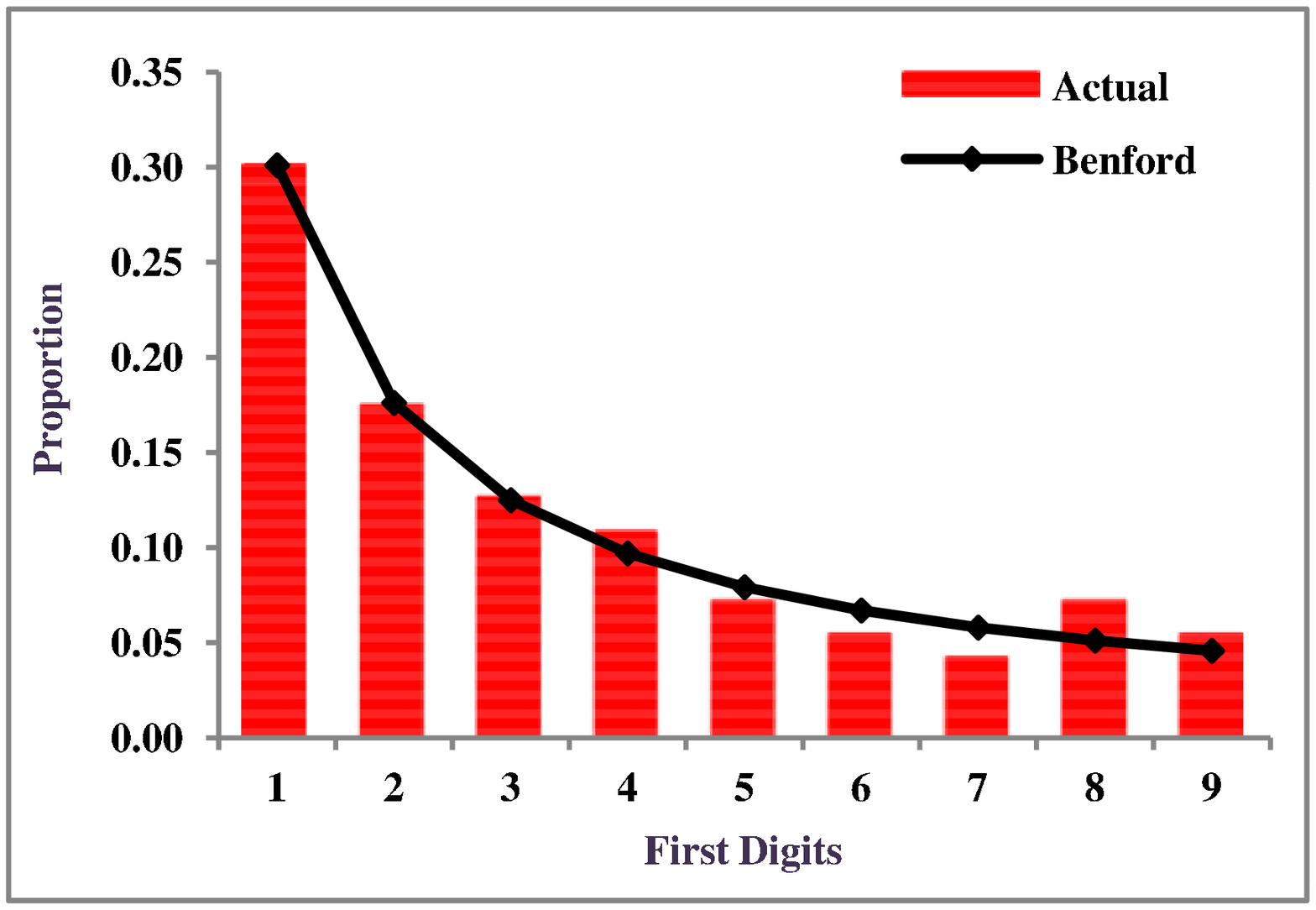}}
\hspace*{-5pt}
\subfigure[New subscriptions]{\label{fig:edge-b}\includegraphics[width=0.55\linewidth, height=0.5\linewidth, clip=]{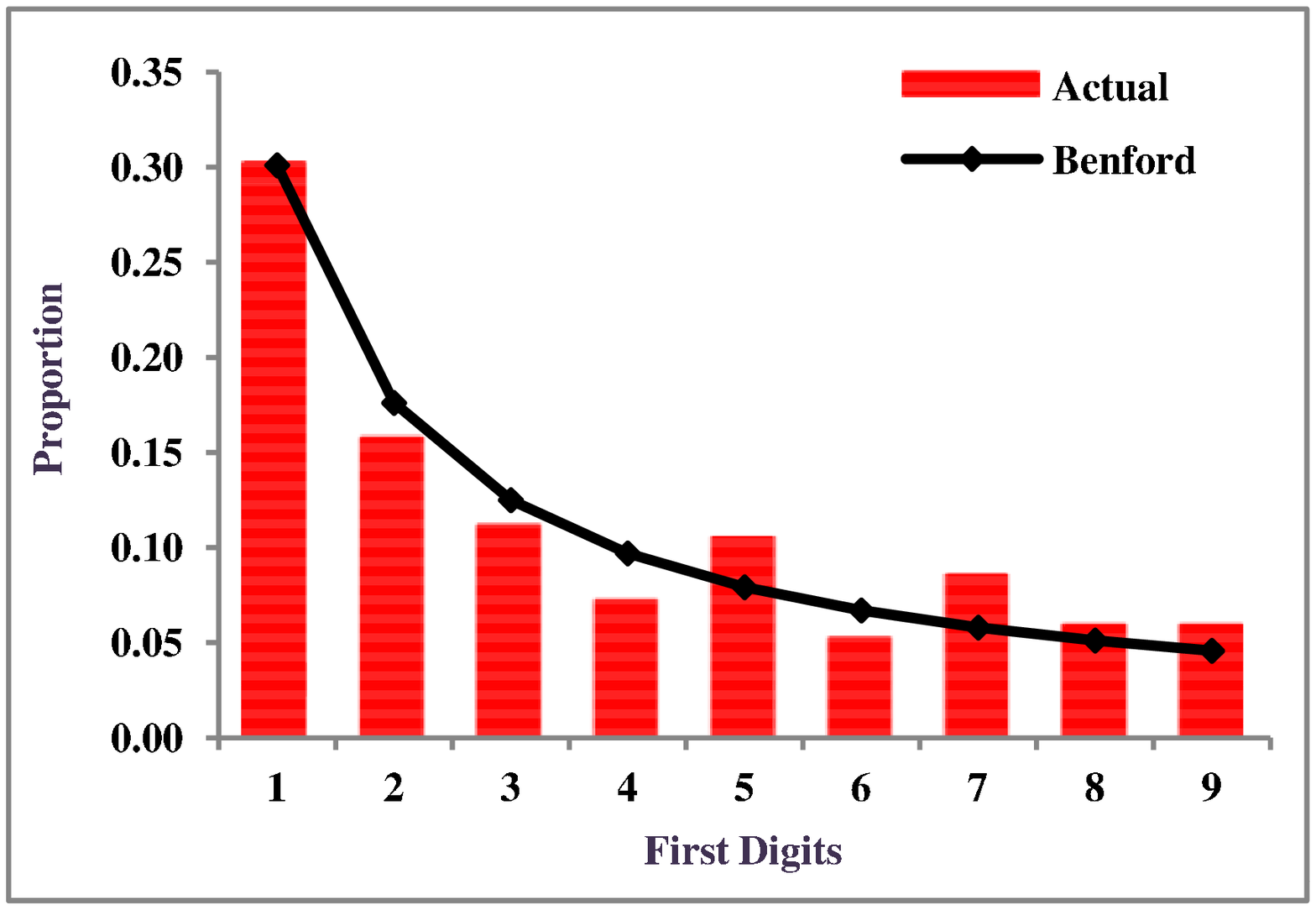}}\\
\hspace*{-50pt}
\vspace*{-100pt}
\subfigure[ Individual magazines]{\label{fig:edge-a}\includegraphics[width=0.55\linewidth, height=0.5\linewidth, clip=]{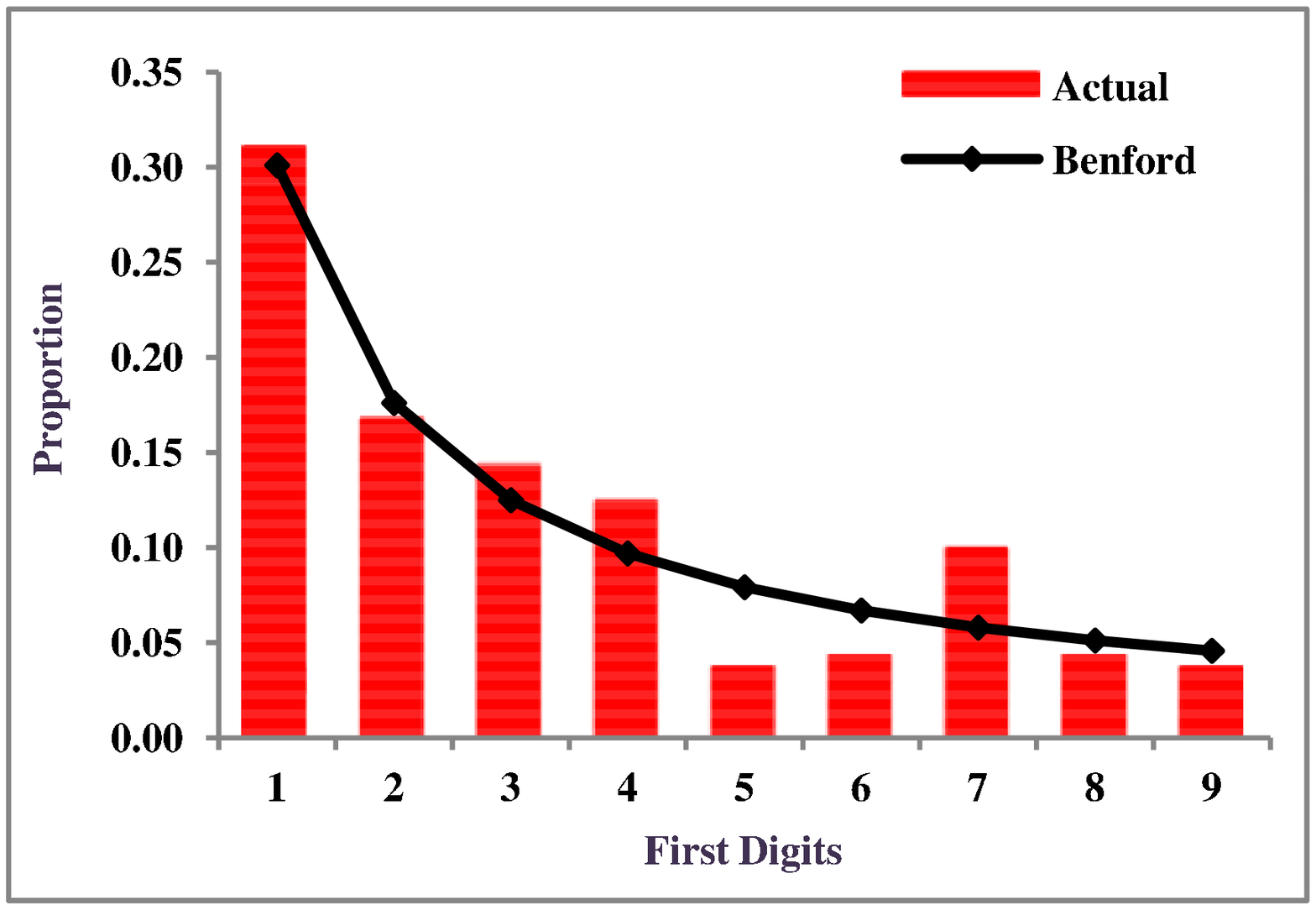}}
\hspace*{-5pt}
\subfigure[Back-calls]{\label{fig:edge-b}\includegraphics[width=0.55\linewidth, height=0.5\linewidth, clip=]{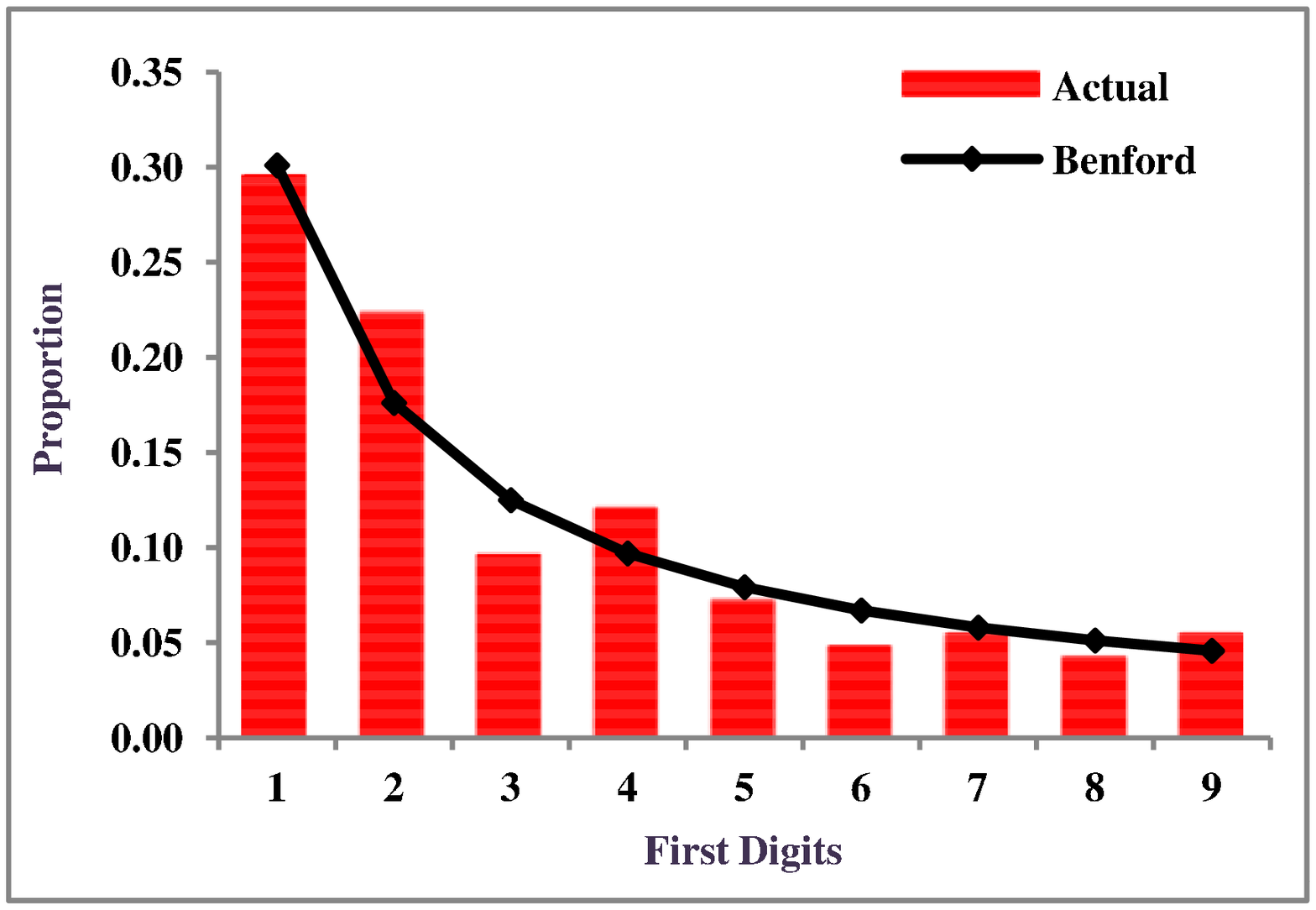}}\\
\hspace*{-30pt}
\vspace*{110pt}
\hspace*{20pt}
\vspace*{-70pt}
\end{tabular}
\vspace*{40pt}
\hspace*{20pt}
\end{minipage}
\end{center}
\caption{Observed and Benford distributions of significant digits for country-wide religious activity data of JW for service report year 1958}
\label{fig:edge}
\end{figure}

\begin{sidewaystable}
%\begin{table}[ph]
\tbl{ The significant digit distribution of country-wise religious activity data of Jehovah's Witnesses}
{\begin{tabular}{@{}lllllllll@{}} \toprule
Service  & Peak publ-  & Baptizat-  &  & Average pion-  & Congregat-  & Hours & Bible & Memorial \\
Report & ishers ($\chi^{2}$) & ions ($\chi^{2}$) &  & eer publishers & ions ($\chi^{2}$) & spent ($\chi^{2}$) & Studies ($\chi^{2}$) & Service\\
Year &  &  & & ($\chi^{2}$) & &  & & Attendance ($\chi^{2}$)\\ \colrule
$2001$ \hphantom{00} & 206  (12.369) & 190 (5.467)&  & 203 (9.406)& 203 (9.717) & 206 (11.701) & 206 (2.564) & 206 (9.323)\\\\
$2002$ \hphantom{00} & 206  (13.990) & 191 (8.944)&  & 202 (8.130)& 204 (13.970) & 206 (9.118) & 206 (4.880) & 206 (9.677)\\\\
$2003$ \hphantom{00} & 207 (11.590) & 190 (10.286) &  & 205 (7.483) & 205 (\bf25.165) & 207 (13.907) & 207 (8.592) & 207 (11.902)\\\\
$2004$ \hphantom{00} & 206 (12.135) & 190 (12.636) &  & 204 (6.608) & 205 (\bf18.503) & 206 (10.237)
 & 206 (13.202) & 206 (10.062)\\\\
$2005$ \hphantom{00} & 207 (10.506) & 190 (8.113) &  & 203 (4.083) & 206 (9.581) & 207 (9.736) & 207 (10.344) & 207 (6.187)\\\\ 
$2006$ \hphantom{00} & 207 (14.073) & 188 (5.385) &  & 203 (3.378) & 206 (9.230) & 207 (11.435) & 207 (10.423) & 207 (15.240)\\\\ 
$2007$ \hphantom{00} & 206 (13.846) & 192 (12.730) &  & 201 (6.297) & 206 (8.416) & 206 (15.074) & 206 (3.794) & 206 (11.074)\\\\
$2008$ \hphantom{00} & 206 (10.460) & 191 (14.853) &  & 202 (11.572) & 205 (\bf16.051) & 206 (14.353) & 204 (6.547) & 206 (9.442)\\\\
$2009$ \hphantom{00} & 206 (12.295) & 193 (11.922) &  & 203 (\bf15.874) & 206 (11.486) & 206 (\bf17.725) & 206 (8.122) & 206 (\bf21.195)\\\\   
$2010$ \hphantom{00} & 206 (13.807) & 194 (2.259) &  & 203 (11.280) & 206 (11.747) & 206 (\bf23.070) & 206 (10.872) & 206 (12.055)\\\\  
$2011$ \hphantom{00} & 206 (\bf16.568) & 191 (2.486) &  & 202 (12.261) & 206 (11.469) & 206 (15.341) & 206 (\bf18.116) & 206 (\bf16.457)\\\\ 
$2012$ \hphantom{00} & 209 (13.917) & 194 (8.792) &  & 207 (11.392) & 209 (12.448) & 209 (10.508)  & 209 (6.009) & 209 (12.496)\\ 
\botrule
\end{tabular} \label{ta1}}
\end{sidewaystable}
%\end{table}

\begin{sidewaystable}
%\begin{table}[ph]
\tbl{ The significant digit distribution of country-wide religious activity data of Jehovah's Witnesses, NR stands for data not reported}
{\begin{tabular}{@{}llllllllllll@{}} \toprule
Service  & Average & Peak publ-  & Total   & Average pion-  & Public   & Individual & Back  & Hours & Bible & New & Number  \\
Report & Publishers & ishers  & Literature  &  eer publishers & Meetings & Magazines & Calls & spent  & Studies & Subscriptions & Companies\\
Year &  &  & ($\chi^{2}$) & ($\chi^{2}$) &  ($\chi^{2}$) & ($\chi^{2}$)  & ($\chi^{2}$) & ($\chi^{2}$) & ($\chi^{2}$) & ($\chi^{2}$) & ($\chi^{2}$) \\ \colrule
$1947$ \hphantom{00} & 88  (9.204) & 88  (5.802) & 86 (11.651) & NR & NR & 80 (3.050) & 87 (15.596) & 88 (13.891) & 84 (7.112) & 77 (13.954) & 82 (5.577)\\\\
$1948$ \hphantom{00} & 96  (12.765) & 96 (9.647) & 93 (14.070) & NR & NR & 91 (9.575) & 94 (11.609) & 94 (10.823) & 92 (7.472) & 88 (14.517) & 90 (5.198)\\\\
$1949$ \hphantom{00} & 104 (8.987) & 104 (10.738) & 102 (12.161) & NR & NR & 100 (15.371) & 103 (9.415) & 103 (6.392) & 101 (5.058) & 96 (5.575) & 96 ( 7.547)
\\\\
$1950$ \hphantom{00} & 115 (10.358) & 115 (5.374) & 113 (3.837) & 99 (3.208) & 94 (5.067) & 106 (5.500) & 115 (5.236) & 
 115 (6.476) & 112 (9.007) & 101 (5.461) & 103 (9.381)
\\\\
$1951$ \hphantom{00} & 121 (6.897) & 120 (0.810) & 117 (\bf21.626) & 108 (4.056) & 106 (15.039) & 115 (9.872) & 119 (9.183) & 119 (9.048) & 117 (4.463) & 109 (\bf21.579) & 105 (11.124)
\\\\ 
$1952$ \hphantom{00} & 117 (9.304) & 117 (8.853) & 115 (3.786) & 105 (8.799) & 99 (11.382) & 113 (6.482) & 116 (9.318) & 117 (12.540) & 115 (9.754) & 107 (\bf18.781) & 101 (5.837)
\\\\ 
$1953$ \hphantom{00} & 140 (4.509) & 140 (2.502) & 138 (2.527) & 120 (4.886) & 122 (8.848) & 137 (7.523)  & 139 (13.758) & 140 (7.131) & 136 (3.884) & 130 (6.608) & 125 (\bf15.932)
\\\\
$1954$ \hphantom{00} & 154 (5.858) & 154 (5.502) & 151 (14.480) & 132 (5.216) & 131 (7.919)  & 149 (5.652) & 152 (11.306) & 154 (11.104) & 149 (8.336) & 141 (6.734) & 134 (\bf16.704)
\\\\
$1955$ \hphantom{00} & 153 (6.186) & 153 (11.545) & 151 (11.380) & 135 (5.546) & 136 (17.709) & 150 (8.965) & 153 (14.490) & 153 (6.912) & 152 (8.466) & 145 (5.347) & 138 (18.507)
\\\\   
$1956$ \hphantom{00} & 157 (4.325) & 155 (4.623) & 157 (3.814) & 142 (7.768) & 140 (7.768) & 154 (5.523) & 157 (4.356) & 157 (9.585 & 156 (14.782) & 150 (4.867) & 143 (\bf21.134)
\\\\  
$1957$ \hphantom{00} & 156 (6.269) & 156 (8.620) &  154 (5.964) & 145 (4.703) & 140 (5.647) & 152 (11.123) & 156 (12.242) & 156 (4.298) & 156 (11.472) & 152 (8.126) & 144 (17.113)
\\\\ 
$1958$ \hphantom{00} & 167 (11.727) & 166 (7.626) & 167 (3.135) & 147 (8.632) & 145 (4.822) & 162 (11.778) & 167 (5.613) & 167 (7.446) & 164 (12.122) & 153 (5.948) & 150 (12.980)\\ 
\botrule
\end{tabular} \label{ta1}}
\end{sidewaystable}
\section{Discussion}
The studies of religious phenomena have mostly focused on the size distribution of various religious groups\cite{Grim1, Hsu}, on modeling the structural changes in the evolution of number of their adherents\cite{Ausloos, Ausloos1} and the factors responsible for driving such changes\cite{Vitanov, Vitanov1, Vitanov2, Abrams}. An important aspect of religious association is that, as part of their religious obligation, members carry out various activities like public preaching, distribution of literature, building of places of worship etc. The scientific studies of religious activities are only fewer. These studies aim to uncover any broad statistical patterns that might exist amongst the data on religious activities such as the growth and decline of the number of churches and temples\cite{Hayward, Hayward1, Ausloos3}, growth and decay in the income and expenses of Antoinist's community in Belgium\cite{Ausloos2}. Further, an analysis of the official Witness data has revealed that the growth dynamics of religious activities follow laws similar to those as in economic activities and scientific research\cite{Picoli}.  
\newline
Prerequisite for any scientific approach to religion is the availability of high quality data. However, the data associated with religious phenomena, being mostly drawn from surveys and opinion polls, is often scarce, limited and subject to caution\cite{Ausloos}. Due to the very controversial nature of the religious census many countries, as a matter of policy, do not collect the data on religious matters\cite{Usreport}. On the other hand, each year religious bodies themselves collect a formidable amount of data on their activities. However, the authenticity of the membership estimates of religious groups become suspicious when striking discrepancies with the results on membership numbers from government censuses are taken into account\cite{Cragun, Johnson}. Being truthful is one of the primary virtues which the members of a religious group must adhere to. Thus religious bodies are less likely to indulge in data misrepresentation as such unethical practices are against the teachings of religions\cite{McGuire, Boone}. Nevertheless, religious movements are often a subject matter of news, gossip, scandals and many a times of government inquiries. Religious associations might tend to exaggerate their membership numbers for political propaganda since social groups that have more members are going to be more attractive for people to join in\cite{Abrams}. Further, given the large number of publishers, in case of Witnesses, which collect and file the reports from individual countries the data is also susceptible to human processing error\cite{Cragun}.
\newline
One statistical tool, routinely used by forensic analysts, to identify possibly manipulated data is the so-called Benford's law. 
This test is based on the supposition that first digits in real data tend to follow the Benford distribution while the digits in tampered and fabricated data do not\cite{Hill2}. Thus significant deviations from the Benford distribution may indicate fraudulent or corrupted data\cite{Nigrini1}. However, one must be cautioned here that any deviation from Benford's law is not a concrete proof of data being corrupt but is only an alert for further evaluating the quality of the data. Nevertheless, Benford's law is useful in analytical procedures for testing the completeness of financial reports\cite{Nigrini1}.
\newline 
Due to their distinctive social and political behavior, the JWs continue to be a subject of intense debate amongst the sociologists of religion.  The official Witness data is indispensable to any models aimed at understanding their massive growth and for making projections of their future membership numbers\cite{Stark}. Hence the quality of the data must be ensured, for any model build on dodgy data is destined to fail. Thus sociologists of religions consider the official data of the group to be an exceptionally precise and credible treasure trove of information and  reassure themselves about the authenticity of the data by arguing that the facts (i) that Witnesses yearly not only report the increase in number of their members but also record the corresponding decrease and (ii) that even the disenchanted ex-members accept these statistics are an indication for the truthfulness of their data\cite{Stark}. A further test of Witness statistics is done by cross-checking against the numbers reported in the censuses of various countries in which the religious identity is asked\cite{Stark, Cragun}. The first two arguments in support of the Witness data are only axiomatic, devoid of any quantitative approach. On the other hand, the comparison with the government censuses for checking the reliability is problematic as the government enumerations of religious groups in itself are questionable. Several countries recognize a particular religion as official religion of the state whereas declare the operations of certain other religious groups as illegal\cite{Usreport}. To strengthen their authority the governments might enhance the membership numbers of state promoted religions whereas suppress the numbers associated with the less favored and banned ones\cite{Johnson}. Thus given the propaganda value the religious groups have and the political incentive governments have in inflating the numbers it is necessary to check the reliability of the data through scientific methods. Testing the conformity of the data against the Benford's law provides an independent method of assessing the reliability of data from religious groups.
\newline
We assessed the reliability of the Witness data using Benford's law and found it to be in general conformity to the law. Our analysis of the records of years 1947 to 1958  (Table 4) and 2000 to 2012 (Table 5) further shows that the conformity to Benford's law has remained consistent over the years. Out of 186 samples (84 in Table 4 and 102 in Table 5 respectively) of data associated with different activities of JWs only 16 samples (10 in Table 4 and 6 in Table 5 respectively) deviate from the law. Such an extent of conformity to Benford's law in spite of the every possibility of human errors creeping in due to the large number of pioneers (Witnesses of high standing), many working underground for fear of persecution in countries where Witnesses are banned, involved in the processing of data is impressive. An earlier study has revealed the presence of statistical patterns similar to those present in economic and research activity amongst the Witness data on total number of publishers, pioneers, total time in hours spent in the public preaching work and monthly average of home courses conducted\cite{Picoli}. We went a step ahead and demonstrated the presence of patterns in the distribution of first digits as predicted by Benford's law amongst the data associated with all the twelve religious activities reported in the yearbooks of JWs.
 
\section{Conclusion}
The technique of testing the data from religious groups against Benford's law is advantageous in being independent of the method of comparison with government censuses estimates which are often controversial. We evaluated, for the first time, the credibility of official JWs data by investigating the country-wide numerical data on the twelve different religious activities of its members for conformity to Benford's law. We found that data on all the religious activities of JWs excellently follow the law.

\section*{Acknowledgments}
The constructive and helpful comments of four anonymous reviewers are highly appreciated. 

%This section should come before the References. Dedications and funding 

%\section*{References}

\end{document}